\documentclass[conference]{IEEEtran}
\usepackage{fancyhdr}
\IEEEoverridecommandlockouts
\usepackage{cite}
\usepackage{amsmath,amssymb,amsfonts}
\usepackage{algorithmicx}
\usepackage{graphicx}
\usepackage{textcomp}
\usepackage{amssymb}
\usepackage{latexsym}
\usepackage{xcolor}
\usepackage{booktabs}
\usepackage{multirow}
\usepackage{balance}

\usepackage{makecell}

\usepackage{url}

\usepackage{amsmath}
\usepackage{bm}
\usepackage{amsthm}
\usepackage{algpseudocode}
\usepackage{algorithmicx,algorithm}

\usepackage{stfloats}
\usepackage{cuted}

\newcommand{\circledx}{\textcircled{x}}
\newcommand{\circledy}{\textcircled{y}}
\newcommand{\circledz}{\textcircled{z}}

\newif\ifshowchanges
\showchangesfalse

\DeclareRobustCommand{\chggg}[1]{%
  \ifshowchanges
    \textcolor{red}{#1}%
  \else
    #1%
  \fi
}

\DeclareRobustCommand{\chg}[1]{%
  \ifshowchanges
    \textcolor{black}{#1}%
  \else
    #1%
  \fi
}

\DeclareRobustCommand{\chgblue}[1]{%
  \ifshowchanges
    \textcolor{black}{#1}%
  \else
    #1%
  \fi
}

\newenvironment{chgblock}
{\ifshowchanges\color{black}\fi}
{}

\newenvironment{chgblockgreen}
{\ifshowchanges\color{black}\fi}
{}

\def\BibTeX{{\rm B\kern-.05em{\sc i\kern-.025em b}\kern-.08em
    T\kern-.1667em\lower.7ex\hbox{E}\kern-.125emX}}
\begin{document}

\title{TokaGLINT: A Scalable GPU-Tailored Implicit Solver for Full 3D Tokamak Electromagnetic Simulations}

\author{
\IEEEauthorblockN{
Zifan Yang\textsuperscript{1,2},
Haoyuan Zhang\textsuperscript{1},
Jialin Li\textsuperscript{3},
Wu Yuan\textsuperscript{1},\\[0.3ex]
Xiazhen Liu\textsuperscript{1},
Jian Zhang\textsuperscript{1,*},
Jianyuan Xiao\textsuperscript{4,*},
Shan Liang\textsuperscript{1,*}
}

\IEEEauthorblockA{
\textsuperscript{1}Computer Network Information Center, Chinese Academy of Sciences, Beijing, China\\
\textsuperscript{2}University of Chinese Academy of Sciences, Beijing, China\\
\textsuperscript{3}Tsinghua University, Beijing, China\\
\textsuperscript{4}University of Science and Technology of China, Hefei, China
}

\thanks{\textsuperscript{*}Corresponding authors
(zhangjian@sccas.cn; xiaojy@ustc.edu.cn;\newline liangshan@sccas.cn).}
}

\maketitle
\thispagestyle{fancy}
\lhead{}
\rhead{}
\chead{}
\lfoot{\footnotesize{
This paper has been accepted at the International Conference for High Performance Computing, Networking, Storage, and Analysis (SC 2026).
}}
\rfoot{}
\cfoot{}
\renewcommand{\headrulewidth}{0pt}
\renewcommand{\footrulewidth}{0pt}

\begin{abstract}
We introduce TokaGLINT, a GPU-accelerated implicit solver for electromagnetic field computations in full 3D tokamak simulations, aimed at efficient large-scale parallel GPU computing. Its central innovation lies in the co-design of hierarchical domain decomposition and a fast exact local solver, where hierarchical partitioning is tailored to match fine-grained intra-card subdomains and exploit the tensor-based solver dedicated to curvilinear-coordinate symplectic CN-FDTD-discretized 3D Maxwell equations. Backed by automated operator fusion and batching customized for the intra-card multi-subdomain structure, the solver decouples unknowns through discrete transformations and leverages tensor-structured computations to achieve high hardware utilization, while preserving the long-time stability characteristic of symplectic discretizations. TokaGLINT scales the electromagnetic field solve beyond 10,000 GPUs, achieving 90.1\% weak and 53.9\% strong scaling efficiency, while delivering a 2.67$\times$ single-node speedup over an unpreconditioned BiCGStab baseline (HIP-enabled HYPRE). It is validated in EAST tokamak simulations within the SymPIC plasma simulation code, enabling high-fidelity long-duration modeling.
\end{abstract}


\section{Introduction}

\chg{
When studying wave heating, current drive, and energetic-particle interactions with the background plasma in tokamaks \cite{fisch1987theory,prater2004heating,fasoli2007physics}, an electromagnetic (EM) fully kinetic model is desirable since it can capture the non-Maxwellian velocity distributions, finite Larmor radius effects, or wave-particle resonances that are central to these processes.
The particle-in-cell (PIC) method is the standard computational approach for simulating EM fully kinetic plasmas. In this method, macro-particles sample the 6D distribution function,} Maxwell's equations govern the evolution of EM fields, which in turn exert Lorentz forces on charged plasma particles to \chg{update} their motion state, while the spatial distribution and motion of plasma particles (i.e., charge and current densities) serve as the source terms of Maxwell's equations, forming a closed \chg{self-consistent} loop.
																			
\begin{figure}[tbp]
\includegraphics[trim=10pt 30pt 10pt 30pt,clip,width=3.2in]{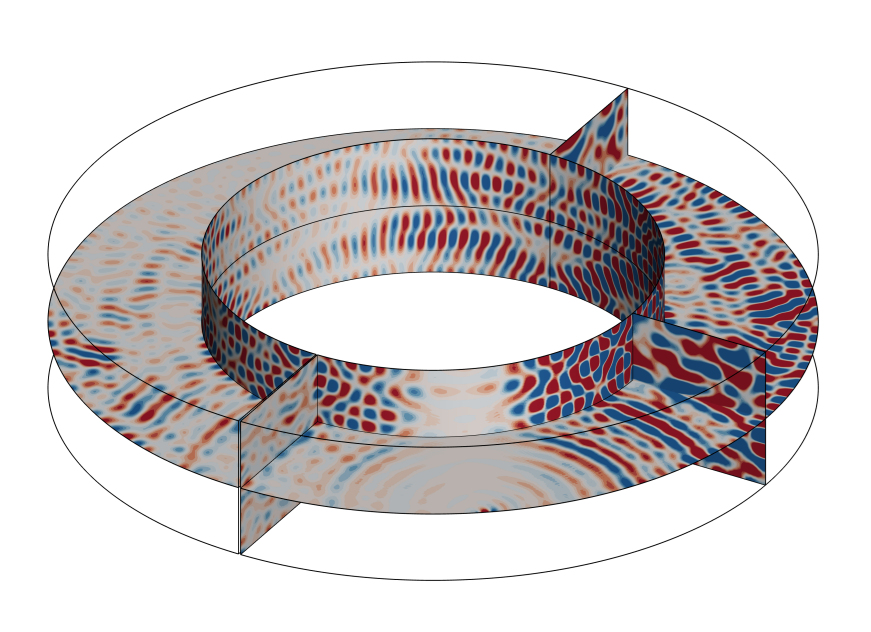}
\caption{3D view of the electric field excited by a point source simulated with the CN-FDTD scheme \chg{accelerated} by TokaGLINT. The simulation domain size and wave frequency are referenced from the lower hybrid wave injection experiment conducted on the EAST Tokamak. }
\label{figwhole}
\end{figure}
\begin{chgblock}
The wave-heating, current-drive, and energetic-particle processes of interest evolve on time scales far longer than the fundamental periods that limit explicit PIC time steps, i.e., the inverse plasma frequency, the inverse cyclotron frequency,
and the EM wave crossing time across a grid cell (the EM Courant-Friedrichs-Lewy). As a result, simulations routinely span millions of time steps. Under these conditions, cumulative discretization errors in physical invariants (energy, charge, or more general, symplectic structure) must remain bounded, a requirement that places strict demands on the numerical scheme.
In response, a family of structure-preserving geometric PIC algorithms has been developed \cite{squire2012geometric,he2015hamiltonian,qin2016canonical,he2016hamiltonian,kraus2017gempic,Morrison2017},
whose space-time discretizations preserve key physical symmetries exactly in a discrete sense.
\end{chgblock}
The SymPIC \cite{sympic2015,sympic2018,sympic2021} \chg{code, a member of this family, features an} explicit symplectic schemes and has been applied to full 3D tokamak simulations \chg{at extreme scale}\cite{sympicSC21}.
\chg{However, explicit time integration is} constrained by the Courant-Friedrichs-Lewy (CFL) condition. \chg{In the low-density plasma regimes targeted here, or in simulations employing a reduced ion-to-electron mass ratio, the EM wave CFL is the binding constraint, impractically} small time steps are required,  making long-time scale simulations computationally inefficient and prohibitively costly.
\begin{chgblock}
Tokamak geometry is toroidal, naturally calling for a cylindrical or toroidal coordinate mesh. SymPIC already operates on a cylindrical mesh, the challenge is to upgrade its EM solver to an implicit formulation that preserves both the symplectic structure and compatibility with the curvilinear discretization.

The Crank-Nicolson (CN) time discretization is the most natural choice for this upgrade. Within SymPIC's discrete action principle, the EM part of the Lagrangian action integral is discretized in time. Applying the mid-point rule to this integral yields,
via the discrete Euler-Lagrange equations, exactly the CN-FDTD formulation
of Maxwell's equations \cite{xu2002unconditionally,sullivan2000electromagnetic,yee1966numerical}. This choice is not unique, other implicit time-integration methods can also remove the EM CFL, but each carries practical drawbacks. The widely used alternating-direction implicit (ADI) FDTD method~\cite{namiki1999adi,zheng2000adi} achieves unconditional stability by splitting each time step into directional substeps, but it sacrifices the discrete symplectic structure and suffers from more severe numerical dispersion than the CN scheme at large time steps~\cite{garcia2002adi,shibayama2006performance}. A symplectic variant of the backward Euler method can be constructed, yet it is only first-order accurate, losing the second-order precision essential for long-time scale integrations. The mid-point/CN scheme is the simplest extension that simultaneously
removes the EM CFL, preserves the symplectic structure exactly,
retains charge conservation, and integrates seamlessly with SymPIC's
existing spatial discretization and current deposition framework, all at the cost of a single linear solve per time step.
\end{chgblock}
In this work, we upgrade SymPIC to the implicit CN-FDTD framework, achieving both high numerical fidelity and improved computational efficiency while retaining excellent long-term stability. \chg{A brief derivation of charge conservation for the field-implicit SymPIC scheme will be given in the supplement material.}

The implementation of CN-FDTD entails the large-scale parallel solution of curl-curl linear systems.
The primary challenge arises from the frequent neighbor and global communication operations inherent in iterative methods for sparse linear systems, which severely hinder parallel scalability.
Domain decomposition preconditioning serves as an effective solution to address this scalability challenge.
A well-designed preconditioner can substantially reduce the number of iterations while lowering the proportion of communication overhead, thereby improving parallel scalability. For the overall performance, the cost associated with preconditioning operations must be offset by the reduction in iteration count and the enhancement in scalability. In general, high-precision preconditioners are more effective in reducing the number of iterations but typically incur higher computational overhead. Thus, there exists a trade-off between the cost and strength of the preconditioner. The exceptional computing power of GPUs offers ample room for the design and optimization of high-strength preconditioners.

Motivated by these facts, we propose TokaGLINT (GPU Linear Iterative Solver with Numerical Transforms for Tokamak), a composite preconditioner based on multi-level domain decomposition and a tensor-form fast solver for subdomains, which is specifically tailored for the large-scale GPU-parallel solution of CN-FDTD. A key feature of this composite preconditioner is its ability to accommodate multiple overlapping subdomains within a single accelerator card. This capability allows us to employ a high-computational-intensity tensor-form solver for each subdomain, while the overlap between subdomains can be efficiently implemented using High Bandwidth Memory (HBM).
When combined with the upper-level traditional overlapping domain decomposition across different cards, this composite preconditioner substantially reduces the number of outer iterative method iterations compared to conventional preconditioners. Specifically, the superior convergence performance of the proposed scheme is quantitatively validated in Section \ref{subsec:overlap_impact}, while its enhanced computational efficiency is demonstrated in Section V.


A further key advantage of the composite preconditioner is that the subdomain solver is specifically designed to preserve the tensor structure of the curl-curl differential operator and the EM field. We develop a fast algorithm that achieves partial decoupling of the unknowns via discrete transformations derived from the differential operator and its \chg{eigenvalue/singular-value} decomposition. While this approach has been demonstrated in the literature \cite{19-nie2008compact,JSC16,CMS16,FlashMP}, extending it to efficiently handle the complex curl-curl operator with variable coefficients introduced by \chg{cylindrical} coordinates poses a highly nontrivial challenge.
Nevertheless, we successfully generalize and adapt this approach to the tokamak simulation regime with substantial and practically impactful speedup, owing not only to the \chg{co-design of} hierarchical domain decomposition framework \chg{ and fast exact subdomain solver} but also to the careful engineering of operator fusion and batching across multiple intra-card subdomains.

Our primary contributions are outlined as follows.
\begin{itemize}
\item {We propose a novel GPU-tailored Hierarchical Additive Schwarz Method (HASM) preconditioner. Built around the co-design of intra-card domain decomposition and physics-driven fast subdomain solvers, this architecture is natively optimized to deliver efficient large-scale parallel GPU solutions for full 3D tokamak EM simulations.}
\item{We develop a fast exact local solver tailored to the linear systems arising from the curvilinear-coordinate symplectic CN-FDTD-discretized 3D Maxwell equations. It leverages discrete transformations to decouple unknowns, exploits tensor-structured operations for accelerated computation, and is seamlessly integrated with the intra-card multi-subdomain framework to enable automated operator fusion and batching, consistently achieving high hardware utilization across diverse simulation scenarios.}
\item{The proposed TokaGLINT
\chggg{delivers outstanding performance, sustaining} 90.1\% weak and 53.9\% strong \chggg{scaling efficiency across more than} 10,000 GPUs, \chggg{together with a} $2.67\times$ \chggg{single-node speedup over an unpreconditioned BiCGStab baseline} (HIP-enabled HYPRE).
Integrated into SymPIC, TokaGLINT has been validated in full-3D EAST tokamak EM simulations, confirming its practicality for real-world fusion \chggg{plasma simulation}.}
\end{itemize}

\section{Related work}
\subsection{EM solver in PIC}
PIC codes are widely recognized as essential tools for plasma simulation, and numerous research teams worldwide have dedicated efforts to advancing this technology. A variety of high-quality full-kinetic PIC software packages have been developed, with representative examples including SMILEI\cite{35-Derouillat2017}, PICADOR\cite{36-Bastrakov2012}, EPOCH\cite{37-Arber2015}, VPIC 2.0\cite{38-Bird2022}, WarpX\cite{41-Fedeli2022}, iPIC3D\cite{42-Markidis2010, 39-Williams2023}, PIConGPU \cite{59-Burau2010, 43-Bussmann2013}, OSIRIS\cite{44-Fonseca2002}, and SymPIC\cite{sympicSC21}. iPIC3D is an implicit PIC code with its field solver running on CPUs.
All other packages adopt explicit numerical schemes and exhibit remarkable scalability when deployed on large-scale computing platforms. For instance, VPIC 2.0 has successfully completed weak scaling tests involving over 17,000 GPUs\cite{38-Bird2022}, while WarpX has been optimized for exascale supercomputers such as Frontier, Fugaku, and Summit, leveraging thousands of AMD and NVIDIA GPUs\cite{41-Fedeli2022}. SymPIC\cite{sympicSC21} scaled up to over 40,000,000 cores on the new Sunway supercomputer.
These explicit PIC codes have been widely applied in plasma physics research, covering areas including laser-plasma interaction, magnetic reconnection, and inertial confinement fusion. Their excellent scalability originates from the local nature of particle updates. However, they are tightly constrained by the CFL condition \chg{when plasma density is low or a reduced ion-electron mass-ratio is used}.

\begin{chgblock}
A complementary strategy is implicit PIC \cite{cohen1982a,langdon1983,chen2011,Chacon2013}, which treats both field and particle advances implicitly. The direct implicit approach~\cite{cohen1982a,langdon1983} linearizes the coupled system about the previous time step, requiring only a single linear solve per time step and avoiding particle-orbit Jacobian evaluations. In contrast, the fully implicit Newton-Krylov approach~\cite{chen2011,Chacon2013} solves the coupled nonlinear system via Newton-Krylov iteration, incurring multiple linear solves and particle-orbit Jacobian evaluations per time step, with irregular communication patterns that severely degrade GPU scalability. Implicit PIC becomes necessary when the plasma-frequency CFL ($\omega_{pe}\Delta t < 2$) is the binding constraint. In the low-density or reduced mass-ratio regimes targeted in this work, however, the EM CFL dominates, and the CN-FDTD field-implicit strategy pursued here captures the majority of the practical speedup at a fraction of the algorithmic complexity.
\end{chgblock}

\subsection{Multilevel domain decomposition}
Domain decomposition methods\cite{keyes1987comparison,keyes1992bridge,17-smith1996domain,22-DDbook2015} are widely recognized as effective solution approaches suitable for parallel computations. Among these techniques, the two-level overlapping Schwarz domain decomposition methods\cite{22-DDbook2015,21-robustDD2011,23-DDML2020} incorporate a coarse space correction mechanism to mitigate the convergence degradation of one-level methods when dealing with a large number of subdomains. Although they exhibit superior convergence speed compared to one-level schemes, they often require an accurate global solution at the coarse level, which ultimately becomes a computational bottleneck. Distinct from the two-level approaches, hierarchical partitioning here in this work serves solely to distinguish between inter-card and intra-card interactions, while the second level is primarily designed to cooperate with batched subdomain solvers based on tensor operations and resolve the data exchange issues among these subdomains.

Concurrent with our research, Ichitaro Yamazaki et al.\cite{46-Yamazaki2023} also adopt the strategy of assigning multiple subdomains to each GPU to improve accelerator utilization. Their approach achieves parallel computation of subdomains on the same GPU through NVIDIA MPS. As noted in \cite{46-Yamazaki2023}, running multiple MPI processes on a single GPU may not be the most optimal approach; however, achieving the same decomposition by having multiple subdomains per MPI process would necessitate substantial innovations in algorithms and extensive software development efforts.
\chg{TokaGLINT follows this latter route: one MPI process manages multiple subdomains on a GPU. This design exposes opportunities for fusing and batching subdomain solves on the same GPU. Detailed discussions are provided in Sec.~IV.B.}

At the subdomain solver level, the family of batched solvers developed by Anzt et al.\cite{47-Anzt2016,48-Anzt2017b,49-Anzt2017c,50-Anzt2017d,51-Anzt2018} is primarily designed for scenarios requiring the batch inversion and factorization of thousands of small matrices, such as the general purpose block-Jacobi and ISAI preconditioning, where the involved matrix dimensions are typically modest ($\leq 100\times100$). In this study, we integrate the batched solver into the overlapping additive Schwarz method (ASM) preconditioning framework, whereas the equivalent matrix size is considerably larger ($ > 3000\times3000$) but the number of matrices is drastically fewer (typically only a few dozen). By exploiting the symmetry of the curl-curl differential operator, we design a discrete transform to partially decouple EM field unknowns, allowing each solver to efficiently handle matrices of this scale.

\subsection{Discrete transform based solver}

A compact representation of the difference operator was first proposed by Frederic in \cite{18-Frederic1973} to achieve storage compression. The authors of \cite{19-nie2008compact} further extended the compact representation to three-dimensional problems via a tensor-based formulation and subsequently proposed fast exact solvers for partial differential equations. The approach has since been generalized to a broader class of PDEs, as demonstrated in \cite{JSC16,CMS16,xGB16}, among others.
Within this line of research, the method was adopted in \cite{FlashMP} for solving complex curl-curl systems in Cartesian coordinates.
\chg{These discrete-transform-based direct solvers achieve high efficiency via unknown decoupling. FlashMP\cite{FlashMP} leverages the symmetry of the curl-curl operator defined on Yee grids to decouple unknowns over separate grid points. However, cylindrical meshes break this symmetry, making FlashMP inapplicable in this setting.}
In the present work, the method is extended to curvilinear coordinates for the first time, and is efficiently coupled with the aforementioned HASM preconditioner, \chg{which adopts the co-design of numerical algorithms and GPU implementations to achieve a favorable balance among convergence, computational complexity and GPU throughput, enabling efficient solving on more than 10,000 GPUs.
}

\section{Algorithm}
\label{sec:algorithm}

\subsection{Background}
\label{subsec:field_implicit_system}
\begin{chgblock}
SymPIC \cite{sympic2015,sympic2018,sympic2021,sympicSC21} constructs explicit charge-conservative symplectic PIC algorithms on cylindrical meshes. In the present work, we revise the temporal discretization of the EM field action integral to a midpoint, equivalently Crank-Nicolson, form. Coupled with SymPIC's existing particle push and charge-conservative current deposition, which remain unchanged from the explicit formulation. This yields a field-implicit, charge-conservative, symplectic PIC scheme on the cylindrical mesh. Charge conservation follows directly from the discrete gauge invariance of the action \cite{sympic2015,sympic2018}, and a brief derivation is provided in the supplement material. The linear system associated with this field-implicit scheme is solved via our TokaGLINT framework.

\end{chgblock}

We start from the EM field equations (\chg{Faraday's and Amp\`ere's laws})
\begin{equation}
\label{eq:classical-maxwell}
\small
\begin{cases}
\displaystyle
\frac{\partial \boldsymbol{E}}{\partial t} = \nabla \times \boldsymbol{B}\chg{-\boldsymbol{J}}, \\[4pt]
\displaystyle
\frac{\partial \boldsymbol{B}}{\partial t} = -\nabla \times \boldsymbol{E},
\end{cases}
\end{equation}
where $\boldsymbol{E}$, $\boldsymbol{B}$, and $\boldsymbol{J}$ denote the
electric field, magnetic field, and electric current density, respectively.

Applying the CN-FDTD scheme to the above system and eliminating $\boldsymbol{B}$ gives the vector-field curl-curl system
\begin{equation}
\left(
\boldsymbol{I}
+
\tfrac{\Delta t^2}{4}
\operatorname{curl}_d^{\mathrm{bw}}
\operatorname{curl}_d^{\mathrm{fw}}
\right)
\boldsymbol{E}^{n+1}
=
\boldsymbol{b},
\label{eq:classical-maxwell-discrete-form}
\end{equation}
where $\boldsymbol{b}$ contains known fields, source terms, boundary contributions, and particle-current contributions. The operators $\operatorname{curl}_d^{\mathrm{fw}}$ and
$\operatorname{curl}_d^{\mathrm{bw}}$ denote the forward and backward
discrete curl operators associated \chg{with staggered discrete field
layout.}

\chg{The SymPIC field representation is geometric: the electric unknown is a
discrete 1-form ~\cite{sympic2018,sympic2021} rather than the vector $\boldsymbol{E}$ itself.}
Accordingly, the vector-field system above must be rewritten in the 1-form unknown used by SymPIC.
The conversion from vector components to 1-form components is determined by the metric of the coordinate system.

For the \chg{tokamak} geometry considered in this work, we adopt \chg{the cylindrical-coordinate
formulation used in SymPIC~\cite{sympic2021,sympicSC21}. The standard cylindrical
coordinates} $(r,\theta,z)$ \chg{are mapped to the straightened coordinates} $(x,y,z)$ as
\begin{equation}
x = r-r_0,\qquad y=r_0\theta,\qquad z=z,
\label{eq:straightened_cylindrical_coord}
\end{equation}
\chg{
where $r_0$ is a fixed radial reference length and the axial coordinate is unchanged.
For typical tokamak applications, $r_0$ is the major radius}, and $\theta$ denotes the toroidal angle.

Under this mapping, the line-element vector \chg{is} $\Delta s = \left(\Delta x,\ \Delta y\,\tfrac{r}{r_0},\ \Delta z\right)$. \chg{It} relates the conventional vector field to the 1-form field
\begin{chgblock}
\begin{equation}
\label{eq:1-form}
\boldsymbol{E}_{\text{1-form}}
=\boldsymbol{P}\boldsymbol{R}\boldsymbol{E}.
\end{equation}

For each point
$c$, define $q_c=r_c/r_0$ and the local diagonal blocks
\begin{equation}
\begin{aligned}
\boldsymbol{P}_c &= \operatorname{diag}(\Delta x,\Delta y,\Delta z),
\qquad
\boldsymbol{R}_c = \operatorname{diag}(1,q_c,1),\\
\boldsymbol{W}_c &= \operatorname{diag}(q_c,1,q_c), \\
\boldsymbol{Q}_c &= \operatorname{diag}(q_c^{-1},q_c,q_c^{-1}) =\boldsymbol{R}_c\boldsymbol{W}_c^{-1}.
\end{aligned}
\label{eq:pointwise-matrices}
\end{equation}
The global diagonal matrices $\boldsymbol{P}$, $\boldsymbol{R}$,
$\boldsymbol{W}$, and $\boldsymbol{Q}$ are assembled from these local
blocks using the point-wise ordering
$(x,y,z)_{c_1},(x,y,z)_{c_2},\ldots$.
\end{chgblock}
The linear system obtained from symplectic CN-FDTD discretization is then given by
\begin{equation}
\small
\boldsymbol{E}^{\,n+1}_{\text{1-form}}
+
\tfrac{\Delta t^2}{4}
\chg{\boldsymbol{P}\boldsymbol{R}}
\operatorname{curl}_d^{\mathrm{bw}}
\operatorname{curl}_d^{\mathrm{fw}}
\left(
\chg{\boldsymbol{R}^{-1}\boldsymbol{P}^{-1}}
\boldsymbol{E}^{\,n+1}_{\text{1-form}}
\right)
=
\chg{\boldsymbol{P}\boldsymbol{R}}\boldsymbol{b}.
\label{eq:symplectic-maxwell-discrete-form}
\end{equation}

\begin{chgblock}

\end{chgblock}

\begin{chgblock}
To expose the tensor-product stencil used by the fast subdomain solver,
we separate the logical-coordinate spacings from the coordinate metric
weights and introduce the Cartesian-like logical-grid curl
$\tilde{\operatorname{curl}}_d$. Its forward form is
\end{chgblock}
\begin{equation}
\footnotesize
\label{eq:new-curl-discrete}
\tilde{\operatorname{curl}}_d^{\mathrm{fw}}\boldsymbol{F}_{i,j,k}
=
\left[
\begin{aligned}
&\frac{\boldsymbol{F}_{z,i,j+1,k}-\boldsymbol{F}_{z,i,j,k}}{\Delta y}
 - \frac{\boldsymbol{F}_{y,i,j,k+1}-\boldsymbol{F}_{y,i,j,k}}{\Delta z} \\[4pt]
&\frac{\boldsymbol{F}_{x,i,j,k+1}-\boldsymbol{F}_{x,i,j,k}}{\Delta z}
 - \frac{\boldsymbol{F}_{z,i+1,j,k}-\boldsymbol{F}_{z,i,j,k}}{\Delta x} \\[4pt]
&\frac{\boldsymbol{F}_{y,i+1,j,k}-\boldsymbol{F}_{y,i,j,k}}{\Delta x}
 - \frac{\boldsymbol{F}_{x,i,j+1,k}-\boldsymbol{F}_{x,i,j,k}}{\Delta y}
\end{aligned}
\right].
\end{equation}

Its relation to the original curvilinear discrete curl $\operatorname{curl}_d$ can be
written as
\begin{equation}
\label{eq:new-curl-and-old-curl}
\tilde{\operatorname{curl}}_d \boldsymbol{F}
=
 \chg{\boldsymbol{W}}
\operatorname{curl}_d
\left(
\boldsymbol{R}^{-1}\boldsymbol{F}
\right).
\end{equation}

\begin{chgblock}
We further introduce the scaled electric unknown
\begin{equation}
\label{eq:Etilde-definition}
\tilde{\boldsymbol{E}}
=
\boldsymbol{P}^{-1}
\boldsymbol{E}_{\text{1-form}} .
\end{equation}

Since $\boldsymbol{P}$, $\boldsymbol{R}$, $\boldsymbol{W}$ and
$\boldsymbol{Q}$ are diagonal under the point-wise ordering, their products and inverses reduce to local component-wise
scalings within each point.
Combining \eqref{eq:new-curl-and-old-curl} with
\eqref{eq:Etilde-definition} and applying straightforward algebra,
\eqref{eq:symplectic-maxwell-discrete-form} becomes
\end{chgblock}
\begin{equation}
\label{eq:symplectic-maxwell-discrete-form-newcurl}
\chg{\tilde{\boldsymbol{E}}^{\,n+1} }
+
\tfrac{\Delta t^2}{4}
\boldsymbol{Q}\,
\tilde{\operatorname{curl}}_d^{\mathrm{bw}}
\left(
\boldsymbol{Q}\,
\tilde{\operatorname{curl}}_d^{\mathrm{fw}}
\chg{\tilde{\boldsymbol{E}}^{\,n+1}}
\right)
=
\chg{\boldsymbol{R}}\boldsymbol{b}.
\end{equation}
\chg{This metric-weighted curl-curl system is the linear system addressed by TokaGLINT.}

\subsection{Pipelined BiCGStab with \chg{H}ASM preconditioning}
\label{sec3b}
\begin{chgblock}
\begin{algorithm}[h]
\caption{\chg{Preconditioned Pipelined BiCGStab}}
\label{PIP}
\footnotesize
\linespread{1.2}\selectfont
\begin{algorithmic}[1]
\begin{chgblock}
\Statex \textbf{Input:} Matrix, RHS, Preconditioner
\Statex \textbf{Output:} Solution
\State Initialization
\Loop
    \State \textbf{computation} local axpby, dot-product partial sums
    \State Begin global sum allreduce 1
    \State \hspace{0.5cm}Apply HASM preconditioner
    \State \hspace{0.5cm}Apply SpMV with halo-exchange overlap
    \State End global sum allreduce 1

    \State \textbf{computation} local axpby, dot-product partial sums
    \State Begin global sum allreduce 2
    \State \hspace{0.5cm}Apply HASM preconditioner
    \State \hspace{0.5cm}Apply SpMV with halo-exchange overlap
    \State End global sum allreduce 2

    \State \textbf{computation} local axpby
\EndLoop
\end{chgblock}
\end{algorithmic}
\end{algorithm}
\end{chgblock}

The outer Krylov subspace iteration in TokaGLINT is realized using a preconditioned pipelined BiCGStab algorithm \cite{pipeline}, whose full procedure is described in Algorithm \ref{PIP}.
In contrast to standard preconditioning setups, we introduce a customized HASM preconditioner. Its key merit lies in a GPU-oriented hierarchical decomposition, which allows a single GPU to handle multiple overlapping subdomains simultaneously, combined with strong, computationally intensive subdomain solvers. \chg{The detailed derivation and formulas of the subdomain solvers are presented in Sec.~III.C, while the hierarchical domain decomposition and its efficient GPU implementation are described in Sec.~IV.}

\chg{In the pipelined BiCGStab framework, the global reductions are overlapped with SpMV and preconditioner applications. This strategy is most effective when the preconditioner application is compute-bound and requires only limited neighbor communication [37], both of which match the design of our HASM preconditioner: the subdomains are compute-intensive, and inter-subdomain communication within a single GPU is realized via HBM. With this framework, the bottleneck caused by global reductions and overlapping-region communications in large-scale parallel execution can be substantially mitigated. This effect is evaluated experimentally in Sec. V.}

\chg{Moreover, the intra-GPU domain decomposition also plays an important role in complexity control and supports operator fusion across subdomains, which we discuss in detail in Sec.~IV.}

\subsection{Fast \chg{subdomain} solver}
\label{subsolve}
We adopt the compact operator notation introduced in \cite{19-nie2008compact}. For any matrix $\boldsymbol{T} = \{T_{ij}\} \in \mathbb{R}^{n \times n}$ and 3D field $\boldsymbol{q} = \{q_{ijk}\} \in \mathbb{R}^{n^3}$, a set of compact operators $\textcircled{x}, \textcircled{y}, \textcircled{z}$ are defined as follows,
\begin {equation}
\label {eq:compact-dot}
\begin {aligned}
\boldsymbol {T} \circledx \boldsymbol {q} &= T_{im} q_{mjk},\\
\boldsymbol {T} \circledy \boldsymbol {q} &= T_{jm} q_{imk},\\
\boldsymbol {T} \circledz \boldsymbol {q} &= T_{km} q_{ijm}.
\end {aligned}
\end {equation}

It is easy to verify that for any $\boldsymbol{T}_1, \boldsymbol{T}_2 \in \mathbb{R}^{n \times n}$,
\begin {equation*}
\begin {aligned}
\boldsymbol{T}_1 \circledx \boldsymbol{T}_2 \circledx \boldsymbol{q} &= (\boldsymbol{T}_1 \boldsymbol{T}_2) \circledx \boldsymbol{q},\\
\boldsymbol{T}_1 \circledx \boldsymbol{q} + \boldsymbol{T}_2 \circledx \boldsymbol{q} &= (\boldsymbol{T}_1 + \boldsymbol{T}_2) \circledx \boldsymbol{q},\\
\boldsymbol{T}_1 \circledx \boldsymbol{T}_2 \circledy \boldsymbol{q} &= \boldsymbol{T}_2 \circledy \boldsymbol{T}_1 \circledx \boldsymbol{q}.
\end {aligned}
\end {equation*}

Recall \eqref{eq:symplectic-maxwell-discrete-form-newcurl} and notice that both curl operators (\ref{eq:new-curl-discrete}) are composed of one-dimensional forward and backward finite differences. The forward difference of field $\boldsymbol{G}$
\begin{equation*}
\boldsymbol{\left(\Delta_x G\right)_{i,j,k}}=\boldsymbol{G_{i+1,j,k}-G_{i,j,k}}
\end{equation*}
can be written in the compact tensor form
\begin{equation*}
\boldsymbol{\Delta_x G}=\boldsymbol{D_{fw}}\circledx \boldsymbol{G},
\end{equation*}
where
\begin{equation*}
\label{eq:define D0}
\vspace{-0.5em}
\small
\boldsymbol{D}_{fw} =
\begin{bmatrix}
-1 & 1  &         &  \\
   & -1 & \ddots  &   \\
   &    & \ddots  & 1 \\
   &    &         &  -1
\end{bmatrix}
 \in \mathbb{R}^{n \times n},
\end{equation*}
and $\boldsymbol{D_{bw}}=\boldsymbol{-D_{fw}^T}$. The backward differences, as well as differences in all other coordinate directions can be represented similarly.

\begin{chgblock}
In \eqref{eq:pointwise-matrices}, $q_c=r_c/r_0$ varies only along the $x$ direction. We define
$\boldsymbol{H}=\operatorname{diag}(q^{-1}_{111}, q^{-1}_{211}, \ldots, q^{-1}_{n_x11})\in \mathbb{R}^{n_x \times n_x}$, where $n_x$, $n_y$, and $n_z$ denote the lengths in the $x$, $y$, and $z$ directions, respectively.
For simplicity, we set $\Delta x=\Delta y=\Delta z=1$. Unless otherwise specified, the 3D field vectors in this subsection are stored in field-major order, whereas Sec.~III.A uses point-major ordering; no additional notation is introduced to distinguish these two orderings. From \eqref{eq:pointwise-matrices} and \eqref{eq:new-curl-discrete}, we then obtain
\begin{equation}
\small
\label{eq:new-curl-discrete-compact-op-form}
\boldsymbol{Q} \tilde{\operatorname{curl}}_d^{\mathrm{fw}}\boldsymbol{E}
=
\left[
\begin{aligned}
& \boldsymbol{H}\circledx \boldsymbol{D_{fw}}\circledy\boldsymbol{e}_z - \boldsymbol{H}\circledx\boldsymbol{D_{fw}}\circledz\boldsymbol{e}_y  \\[4pt]
& \boldsymbol{H}^{-1}\circledx\boldsymbol{D_{fw}}\circledz\boldsymbol{e}_x - \boldsymbol{H}^{-1}\circledx\boldsymbol{D_{fw}}\circledx\boldsymbol{e}_z  \\[4pt]
& \boldsymbol{H}\circledx\boldsymbol{D_{fw}}\circledx\boldsymbol{e}_y - \boldsymbol{H}\circledx\boldsymbol{D_{fw}}\circledy\boldsymbol{e}_x
\end{aligned}
\right],
\end{equation}
\end{chgblock}
\begin{chgblock}
where $\boldsymbol{E}=(\boldsymbol{e}_x,\boldsymbol{e}_y,\boldsymbol{e}_z)^T$.

This gives the alternative form \eqref{eq:maxwell-compact-form} of \eqref{eq:symplectic-maxwell-discrete-form-newcurl}, where the unknown is $\boldsymbol{E}$. For simplicity, both sides have been multiplied by $\beta=4/\Delta t^2$, and the resulting scalar factor and the right-hand-side matrix $\boldsymbol{R}$ have been absorbed into $\boldsymbol{b}=(\boldsymbol{b}_x,\boldsymbol{b}_y,\boldsymbol{b}_z)^T$.
\end{chgblock}
Let $\boldsymbol{U,S,V}$ be the \chg{singular value decomposition} of $\boldsymbol{D_{fw}}$,
\begin{equation*}
\label{eq:Dfw_Dbw}
\boldsymbol{D_{fw}}=\boldsymbol{USV^T} \mbox{\hspace{2mm}and\hspace{2mm}}\boldsymbol{D_{bw}}=\boldsymbol{-VSU^T}.
\end{equation*}
Exploring the symmetry of the double-curl system, we can define discrete transforms in tensor form to decouple the unknowns in ($\ref{eq:symplectic-maxwell-discrete-form-newcurl}$). However, in cylindrical coordinates, this decoupling is only partial, as the metric $\boldsymbol{Q}$ appears between the two curl operators.
Define a forward transformation (mapping the electric field vector and right-hand side vector to the feature space)
\vspace{-0.5em}
\begin{equation}
 \label{eq:transform}
 \begin{aligned}
\boldsymbol{h}_x &= \boldsymbol{V}^\mathrm{T}\circledy  \boldsymbol{V}^\mathrm{T}\circledz\boldsymbol{b}_x \\
\boldsymbol{h}_y &= \boldsymbol{U}^\mathrm{T}\circledy  \boldsymbol{V}^\mathrm{T}\circledz\boldsymbol{b}_y \\
 \boldsymbol{h}_z &= \boldsymbol{V}^\mathrm{T}\circledy  \boldsymbol{U}^\mathrm{T}\circledz\boldsymbol{b}_z
 \end{aligned}
 \vspace{-0.5em}
\end{equation}

The corresponding backward transformation (mapping from the feature space back to the physical space) is defined as:
\begin{equation}
 \label{eq:inv transform}
 \begin{aligned}
\boldsymbol{e}_x & =\boldsymbol{V}\circledy  \boldsymbol{V}\circledz\boldsymbol{f}_x \\
\boldsymbol{e}_y & =\boldsymbol{U}\circledy  \boldsymbol{V}\circledz\boldsymbol{f}_y \\
\boldsymbol{e}_z & =\boldsymbol{V}\circledy  \boldsymbol{U}\circledz\boldsymbol{f}_z
 \end{aligned}
 \vspace{-0.5em}
\end{equation}
where $\boldsymbol{f}_x,\boldsymbol{f}_y,\boldsymbol{f}_z$ are the vectors to be solved in the feature space. The transforms essentially correspond to dense matrix-matrix multiplications in the size of $n_i\times n_i$ and $n_i\times(n_jn_k)$.
\vspace{-3em}

\begin{strip}
\vspace{-1em}
\begin{chgblock}
\begin{equation}
\begin{aligned}
\left(\beta\boldsymbol{I}
-\boldsymbol{H}^2 \circledx \boldsymbol{D}_{bw}\boldsymbol{D}_{fw} \circledy
-\boldsymbol{D}_{bw}\boldsymbol{D}_{fw} \circledz
\right)\boldsymbol{e}_x
+\boldsymbol{H}^2\boldsymbol{D}_{fw} \circledx \boldsymbol{D}_{bw} \circledy \boldsymbol{e}_y
+\boldsymbol{D}_{bw} \circledz \boldsymbol{D}_{fw} \circledx \boldsymbol{e}_z
= \boldsymbol{b}_x, \\
\boldsymbol{H}^{-1}\boldsymbol{D}_{bw}\boldsymbol{H} \circledx \boldsymbol{D}_{fw} \circledy \boldsymbol{e}_x
+\left(
\beta\boldsymbol{I}
-\boldsymbol{D}_{bw}\boldsymbol{D}_{fw} \circledz
-\boldsymbol{H}^{-1}\boldsymbol{D}_{bw}\boldsymbol{H}\boldsymbol{D}_{fw} \circledx
\right) \boldsymbol{e}_y
+\boldsymbol{D}_{bw} \circledz \boldsymbol{D}_{fw} \circledy \boldsymbol{e}_z
=\boldsymbol{b}_y, \\
\boldsymbol{H}\boldsymbol{D}_{bw}\boldsymbol{H}^{-1} \circledx \boldsymbol{D}_{fw} \circledz \boldsymbol{e}_x
+\boldsymbol{H}^2\circledx \boldsymbol{D}_{bw} \circledy \boldsymbol{D}_{fw} \circledz \boldsymbol{e}_y
+\left(
 \beta\boldsymbol{I}
-\boldsymbol{H}\boldsymbol{D}_{bw}\boldsymbol{H}^{-1} \boldsymbol{D}_{fw} \circledx
-\boldsymbol{H}^2 \circledx \boldsymbol{D}_{bw}\boldsymbol{D}_{fw} \circledy
\right)\boldsymbol{e}_z
= \boldsymbol{b}_z.
\end{aligned}
\label{eq:maxwell-compact-form}
\end{equation}
\end{chgblock}
\vspace{-1em}
\end{strip}

\chg{We first illustrate the forward transformation using the $\boldsymbol{e}_y$ term in the first equation of \eqref{eq:maxwell-compact-form} as an example. This term is $\boldsymbol{H}^2\boldsymbol{D}_{fw} \circledx \boldsymbol{D}_{bw} \circledy \boldsymbol{e}_y$. According to \eqref{eq:transform}, the corresponding forward transform operator is $\boldsymbol{V}^{\mathrm{T}}\circledy \boldsymbol{V}^{\mathrm{T}}\circledz$ for the first equation. This simplifies to}

\begin{figure}[!t]
  \centering
\includegraphics[width=3in]{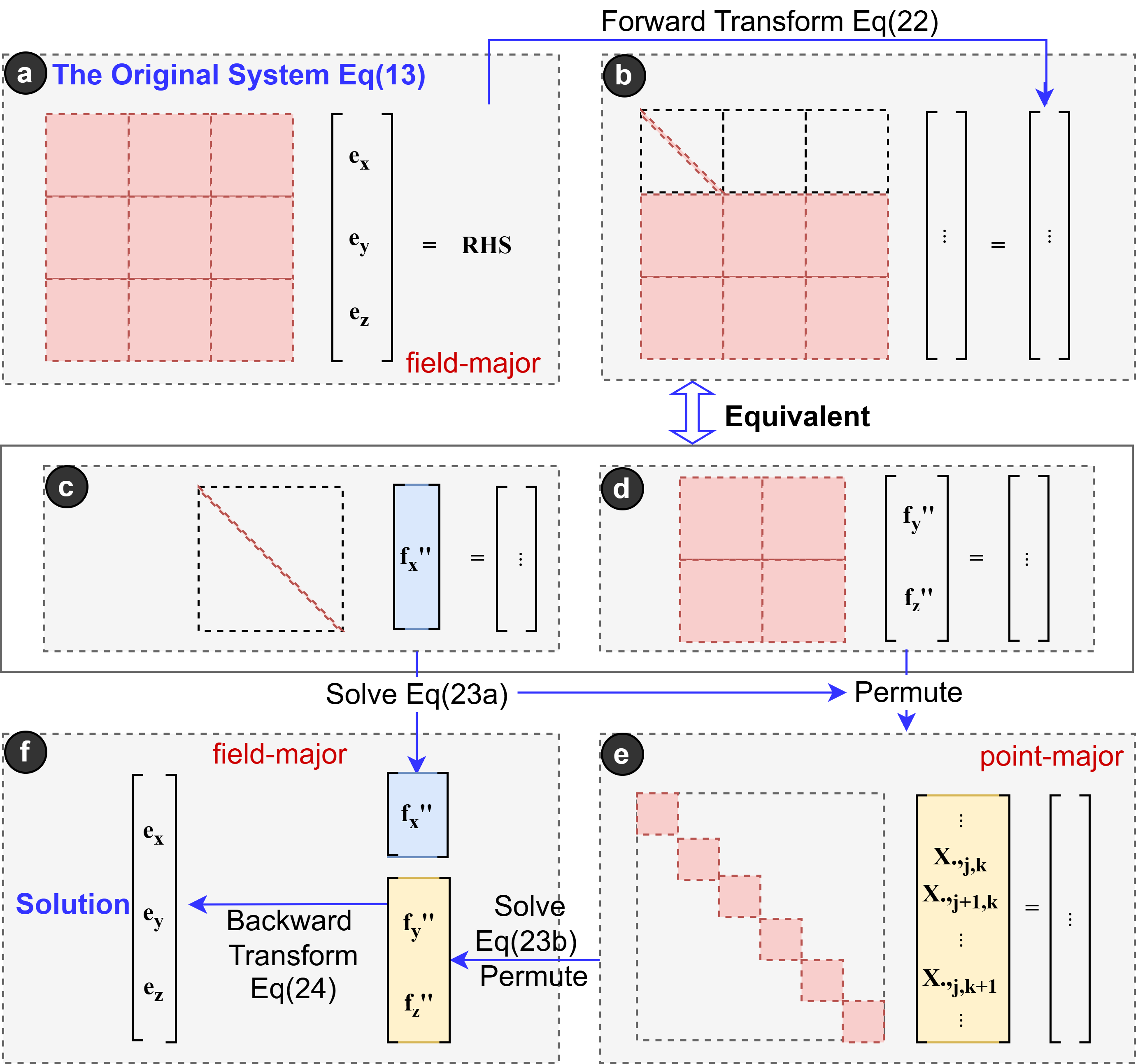}
\caption{Fast exact subdomain solver.
\chg{The original physical-space system (a) is transformed into the feature-space system (b), where the discrete transforms partially decouple the unknowns.
Using the block structure, the system is algebraically reduced to two equivalent subsystems: a diagonal subsystem (c) and a two-component subsystem coupled along $x$-lines (d).
After permuting (d) from field-major to point-major ordering, the coupled subsystem becomes block diagonal (e).
Solving (c) and (e), followed by recombination and the backward transform, gives the solution (f) of the original system.}}
  \label{fig:subsolve}
\vspace{-1.8em}
\end{figure}

\begin{chgblock}
\begin{equation}
 \label{eq:transform on ey}
 \begin{aligned}
 & \boldsymbol{V}^\mathrm{T}\circledy  \boldsymbol{V}^\mathrm{T}\circledz\boldsymbol{H}^2\boldsymbol{D}_{fw} \circledx \boldsymbol{D}_{bw} \circledy \boldsymbol{e}_y \\
 =&-\boldsymbol{V}^\mathrm{T}\circledy  \boldsymbol{V}^\mathrm{T}\circledz\boldsymbol{H}^2 \boldsymbol{D_{fw}}  \circledx \boldsymbol{VSU^T} \circledy\boldsymbol{U}\circledy  \boldsymbol{V}\circledz\boldsymbol{f}_y \\
 =&-\boldsymbol{H}^2 \boldsymbol{D_{fw}}  \circledx \boldsymbol{S} \circledy\boldsymbol{f}_y.
 \end{aligned}
\end{equation}
\end{chgblock}

\chg{By applying the same derivation and simplification to all terms, the full system \eqref{eq:maxwell-compact-form} is transformed into}
\begin{equation}
\label{cyeq}
\begin{pmatrix}
\begin{array}{ccc}
& &\\
& \boldsymbol{M} &\\
& &\\
\end{array}
\end{pmatrix}
\begin{pmatrix}
\boldsymbol{f_x} \\
\boldsymbol{f_y} \\
\boldsymbol{f_z}
\end{pmatrix}
=
\begin{pmatrix}
\boldsymbol{h_x} \\
\boldsymbol{h_y} \\
\boldsymbol{h_z}
\end{pmatrix},
\end{equation}
where $\boldsymbol{M}$ takes the following form
\begin{equation*}
\tiny
\setlength{\arraycolsep}{2pt}
\renewcommand{\arraystretch}{0.9}
\left(\!
\begin{array}{ccc}
\beta I + H^2\circledx S^2\circledy + S^2\circledz  & -H^2D_{fw}\circledx S\circledy & -D_{fw}\circledx S\circledz \\
H_1\circledx S\circledy& \beta I + S^2\circledz - H_1 D_{fw}\circledx & -S \circledy S\circledz \\
H_2\circledx S\circledz & -H^2\circledx S\circledy S\circledz & \beta I-H_2D_{fw}\circledx+H^2\circledx S^2\circledy \\
\end{array}
\!\right)
\end{equation*}
and $\boldsymbol{H_1}=\boldsymbol{H^{-1}D_{bw}H}$, $\boldsymbol{H_2}=\boldsymbol{HD_{bw}H^{-1}}$. \chg{Note that $\boldsymbol{H}$ and $\boldsymbol{H^{-1}}$ are applied only to the $\circledx$ direction; If $n_x$, $n_y$, and $n_z$ differ, $\boldsymbol{D}_{fw}$ and $\boldsymbol{D}_{bw}$ should be replaced by direction-dependent matrices of the corresponding sizes, with $\boldsymbol{U}$, $\boldsymbol{S}$ and $\boldsymbol{V}$ adjusted accordingly.}

\chg{Per Definition (\ref{eq:compact-dot}), the compact operator only generates node dependencies on the field  $\boldsymbol{q}$ along the operating direction. When diagonal matrices $\boldsymbol{T}$ are applied, it introduces no cross-node dependencies.}
Notice that in $\boldsymbol{M}$, the linear system in the feature space, only the diagonal singular value matrix $\boldsymbol{S}$ is applied to the $\circledy$ and $\circledz$ directions.
\chg{From the above observation, we can easily verify that the system is decoupled in these two directions.}
\begin{chgblockgreen}
For example, expanding the first block row in \eqref{cyeq} at point (i, j, k) yields
\begin{equation*}
\begin{array}{c}
\left(\beta+s_j^2\frac{r_0^2}{r_i^2}+s_k^2\right)f_{x,i,j,k}
-s_j\frac{r_0^2}{r_{i}^2}f_{y,i+1,j,k} +s_j\frac{r_0^2}{r_i^2}f_{y,i,j,k}  \\
-s_kf_{z,i+1,j,k} +s_kf_{z,i,j,k}
= h_{x,i,j,k}.
\end{array}
\end{equation*}
\end{chgblockgreen}
The coupling in the remaining direction is represented by the terms involving $\boldsymbol{H}_1$, $\boldsymbol{H}_2$ \chg{ and $D_{fw}$}.

\chg{Since $\boldsymbol{H}$ is diagonal,} the upper-left block of the field-major matrix $\boldsymbol{M}$, namely $\beta \boldsymbol{I} + \boldsymbol{H}^2\circledx \boldsymbol{S}^2\circledy + \boldsymbol{S}^2\circledz$, is diagonal.
\chg{Representing $\boldsymbol{M}$ in block-matrix form, \eqref{cyeq} becomes}
\begin{chgblock}
\begin{equation}
\label{eq:M-in-block-form}
\begin{pmatrix}
\begin{array}{ccc}
\boldsymbol{D}_{11} & \boldsymbol{A}_{12}&\boldsymbol{A}_{13} \\
\boldsymbol{A}_{21} & \boldsymbol{A}_{22}&\boldsymbol{A}_{23} \\
\boldsymbol{A}_{31} & \boldsymbol{A}_{32}&\boldsymbol{A}_{33}
\end{array}
\end{pmatrix}
\begin{pmatrix}
\boldsymbol{f_x} \\
\boldsymbol{f_y} \\
\boldsymbol{f_z}
\end{pmatrix}
=
\begin{pmatrix}
\boldsymbol{h_x} \\
\boldsymbol{h_y} \\
\boldsymbol{h_z}
\end{pmatrix}.
\end{equation}
We introduce the following auxiliary variables:
\begin{equation}
 \label{eq:define newf}
 \begin{aligned}
\boldsymbol{f'}_x &=\boldsymbol{f}_x, \\
\boldsymbol{f'}_y &= \boldsymbol{D}_{11}^{-1}\boldsymbol{A}_{12}\boldsymbol{f}_y, \\
\boldsymbol{f'}_z &= \boldsymbol{D}_{11}^{-1}\boldsymbol{A}_{13}\boldsymbol{f}_z.
 \end{aligned}
\end{equation}
Then \eqref{eq:M-in-block-form} becomes
\begin{equation}
\label{eq:M-in-block-form2}
\small
\begin{pmatrix}
\boldsymbol{D}_{11} & \boldsymbol{D}_{11}&\boldsymbol{D}_{11} \\
\boldsymbol{A}_{21} & \boldsymbol{A}_{22}\boldsymbol{A}_{12}^{-1}\boldsymbol{D}_{11}&\boldsymbol{A}_{23}\boldsymbol{A}_{13}^{-1} \boldsymbol{D}_{11}\\
\boldsymbol{A}_{31} & \boldsymbol{A}_{32}\boldsymbol{A}_{12}^{-1}\boldsymbol{D}_{11}&\boldsymbol{A}_{33}\boldsymbol{A}_{13}^{-1}\boldsymbol{D}_{11}
\end{pmatrix}
\begin{pmatrix}
\boldsymbol{f'_x} \\
\boldsymbol{f'_y} \\
\boldsymbol{f'_z}
\end{pmatrix}
=
\begin{pmatrix}
\boldsymbol{h_x} \\
\boldsymbol{h_y} \\
\boldsymbol{h_z}
\end{pmatrix}.
\end{equation}
Next, we introduce other auxiliary variables:
\begin{equation}
 \label{eq:define newf2}
 \begin{aligned}
\boldsymbol{f''}_x &=\boldsymbol{f'_x}+\boldsymbol{f'_y}+\boldsymbol{f'_z}, \\
\boldsymbol{f''}_y &= \boldsymbol{D}_{11} \boldsymbol{f'}_y , \\
\boldsymbol{f''}_z &= \boldsymbol{D}_{11} \boldsymbol{f'}_z .
 \end{aligned}
\end{equation}
and new forward transformation
\begin{equation}
 \label{eq:newtransform}
 \begin{aligned}
\boldsymbol{h'}_x &= \boldsymbol{h}_x = \boldsymbol{V}^\mathrm{T}\circledy  \boldsymbol{V}^\mathrm{T}\circledz\boldsymbol{b}_x, \\
\boldsymbol{h'}_y &= \boldsymbol{A}_{21}^{-1}\boldsymbol{h_y}
= \boldsymbol{H}_1^{-1}\circledx  \boldsymbol{S}^{-1} \boldsymbol{U}^\mathrm{T}\circledy \boldsymbol{V}^\mathrm{T}\circledz \boldsymbol{b}_y, \\
 \boldsymbol{h'}_z &=\boldsymbol{A}_{31}^{-1}\boldsymbol{h_z}
= \boldsymbol{H}_2^{-1}\circledx  \boldsymbol{V}^\mathrm{T}\circledy \boldsymbol{S}^{-1} \boldsymbol{U}^\mathrm{T}\circledz \boldsymbol{b}_z.
 \end{aligned}
\end{equation}
Then, by left-multiplying the second and third equations in \eqref{eq:M-in-block-form2} by $\boldsymbol{A}_{21}^{-1}$ and $\boldsymbol{A}_{31}^{-1}$, respectively, we obtain
\begin{subequations}
\label{eq:M-in-block-form3}
\small
\begin{align}
&\boldsymbol{D}_{11}\boldsymbol{f''}_x =\boldsymbol{h'}_x, \label{eq:M-in-block-form3-1}\\
&
\begin{pmatrix}
\begin{array}{ccc}
 \boldsymbol{A}_{21}^{-1}\boldsymbol{A}_{22}\boldsymbol{A}_{12}^{-1}-\boldsymbol{D}_{11}^{-1}& \boldsymbol{A}_{21}^{-1}\boldsymbol{A}_{23}\boldsymbol{A}_{13}^{-1}-\boldsymbol{D}_{11}^{-1}\\
\boldsymbol{A}_{31}^{-1}\boldsymbol{A}_{32}\boldsymbol{A}_{12}^{-1}-\boldsymbol{D}_{11}^{-1}& \boldsymbol{A}_{31}^{-1}\boldsymbol{A}_{33}\boldsymbol{A}_{13}^{-1}-\boldsymbol{D}_{11}^{-1}
\end{array}
\end{pmatrix}
\begin{pmatrix}
\boldsymbol{f''_y} \\
\boldsymbol{f''_z}
\end{pmatrix}  \nonumber\\
&=
\begin{pmatrix}
\boldsymbol{h'_y} - \boldsymbol{f_x''} \\
\boldsymbol{h'_z} - \boldsymbol{f_x''}
\end{pmatrix}. \label{eq:M-in-block-form3-2}
\end{align}
\end{subequations}
\end{chgblock}

\begin{chgblock}
Since $\boldsymbol{D}_{11}$ is diagonal, $\boldsymbol{f''}_{x}$ in \eqref{eq:M-in-block-form3-1} can be solved directly. Moreover, all $\boldsymbol{A}_{*}$ matrices and their inverses are decoupled in the $y$ and $z$ directions. Therefore, after permutation, the $2\times2$ block matrix of \eqref{eq:M-in-block-form3-2} becomes a block-diagonal matrix with $n_y n_z$ dense blocks, each of size $(2n_x)\times(2n_x)$. By precomputing the inverses of these $(2n_x)\times(2n_x)$ blocks during initialization, $\boldsymbol{f''}_{y}$ and $\boldsymbol{f''}_{z}$ can be obtained efficiently. The corresponding storage cost is $4n_x^2 n_y n_z$, which scales as $O(N^{4/3})$ for balanced subdomains with $n_x\sim n_y\sim n_z$ and $N=n_x n_y n_z$.

By combining \eqref{eq:inv transform}, \eqref{eq:define newf}, and \eqref{eq:define newf2} with the definitions of $\boldsymbol{A}_{12}$ and $\boldsymbol{A}_{13}$, the solution of the linear system is obtained through \eqref{eq:new inv transform}. The composed $\circledx$, $\circledy$, and $\circledz$ operations in this expression define the new backward transformation.
\begin{equation}
 \label{eq:new inv transform}
 \begin{aligned}
\boldsymbol{e}_x & =\boldsymbol{V}\circledy  \boldsymbol{V}\circledz \left( \boldsymbol{f''}_x-\boldsymbol{D}_{11}^{-1}\boldsymbol{f''}_y-\boldsymbol{D}_{11}^{-1}\boldsymbol{f''}_z \right), \\
\boldsymbol{e}_y & =-\boldsymbol{D}_{fw}^{-1}\boldsymbol{H}^{-2}\circledx \boldsymbol{U}\boldsymbol{S}^{-1}\circledy  \boldsymbol{V}\circledz\boldsymbol{f''}_y, \\
\boldsymbol{e}_z & =-\boldsymbol{D}_{fw}^{-1}\circledx \boldsymbol{V}\circledy  \boldsymbol{U}\boldsymbol{S}^{-1}\circledz\boldsymbol{f''}_z.
 \end{aligned}
\end{equation}
\end{chgblock}

\chg{Figure~\ref{fig:subsolve} illustrates the workflow of this fast exact subdomain solver. }

\subsection{Preconditioning effect}
\label{subsec:overlap_impact}
\chg{In the following tables, L1 and L2 denote the first-level inter-GPU and second-level intra-GPU ASM subdomains, respectively. The notation ``count'' gives the number of subdomains, and ``size'' gives the grid size of each subdomain; both are written as $a\times b\times c$ in the $x$, $y$, and $z$ directions. The parameter $l$ denotes the overlap width used in HASM.}
\begin{table}[!htbp]
  \centering
  \caption{Comparison of BiCGStab Iterative Convergence Steps Under Different Preconditioning Settings. Parameters:$tol_r=10^{-12}$, $\Delta x=1.1$, $\Delta y=1.4$, $\Delta z = 1.0$, $r_0=192$, \chg{L1 count=$2\times2\times2$, L1 size $=32\times32\times32$; TokaGLINT: L2 count=$2\times2\times2$}. MAX: Reached maximum iterations without convergence. DIV: Solution diverged during computation.}
  \label{tab:precond-test}
  \begin{tabular}{lccccc}
\toprule
     $\Delta t$ &1.0 & 2.0  & 4.0 & 8.0& 16.0  \\
\midrule
  Baseline NON\_PRE & 16   & 34   & 60   & 102   & 167 \\
\midrule
    JACOBI (PETSc) &16    &  34  & 61   &  96  & 174 \\
    SOR (PETSc)& 12   &25    &  51  & 82   & 148 \\
    ISAI (HYPRE) & 41   & MAX    & 24   & 46   & 100 \\
    ILUT (HYPRE) & 15   & 31   & 115   & MAX   & DIV \\
    AMG (HYPRE)  & 5   & 9   & 18   &  37  & 63 \\
     \chg{TokaGLINT's HASM($l=1$)}& \chg{3} & \chg{5} & \chg{8} & \chg{12} & \chg{19} \\
    \chg{TokaGLINT's HASM($l=2$)}& \chg{2} & \chg{3} & \chg{6} & \chg{9} & \chg{14} \\
    TokaGLINT's HASM($l=3$)& 2   & 3   & 4   & 7   & 11 \\
    \chg{TokaGLINT's HASM($l=4$)}& \chg{2} & \chg{2} & \chg{4} & \chg{7} & \chg{10} \\
\bottomrule
  \end{tabular}
\end{table}

To assess the extent to which preconditioning improves convergence speed, we conduct comparative evaluations against conventional preconditioning techniques. All general-purpose preconditioners are realized through interfaces to the PETSc and HYPRE libraries\chg{, with their default parameter settings}.
\chg{Table~\ref{tab:precond-test} reports the BiCGStab iteration counts under different $\Delta t$. TokaGLINT's HASM requires fewer iterations than the general-purpose preconditioners, especially for large time steps.  The results also show improved convergence as the overlap size increases. However, enlarging the overlap region also raises computational overhead and data movement costs (both inter-GPU and intra-GPU). In terms of overall runtime, this creates a trade-off between convergence and per-iteration cost. In most cases, an overlap size of 3 delivers the optimal overall performance.}

\chgblue{Table~\ref{tab:precond-asm} reports the weak-scaling convergence as the total problem size grows proportionally with the number of accelerators.}  As the parallel scale is increased, the number of BiCGStab iterations remains nearly constant at approximately 7-8 steps.
The results demonstrate robust and scalable convergence behavior for large-scale parallel simulations, which provides a solid basis for applying the proposed method to ultra-large-scale EM problems.
\begin{table}[htbp]
  \centering
  \caption{\chg{Weak-scaling convergence of TokaGLINT. Parameters: $\mathrm{tol}_r=10^{-12}$, $\Delta x=1.1$, $\Delta y=1.4$, $\Delta z=1.0$, $\Delta t=8.0$, $r_0=1920$, L1 size $=128\times128\times128$, $l=3$, L2 count $=4\times4\times4$. Accelerators equals the product of the L1 count.}}
  \label{tab:precond-asm}
  \small
  \begin{tabular}{@{}cccc@{}}
    \toprule
    Accelerators & \chg{L1 count} &    BiCGStab Steps \\
    \midrule
    8 & \chg{$2\times2\times2$} & 7.40\\
    64 & \chg{$4\times4\times4$} &  7.90 \\
    512 & \chg{$8\times8\times8$} &7.75  \\
    4096 & \chg{$16\times16\times16$}    & 8.00 \\
    \bottomrule
  \end{tabular}
\end{table}

\begin{figure*}[htbp!]
  \centering
\includegraphics[width=6.5in]{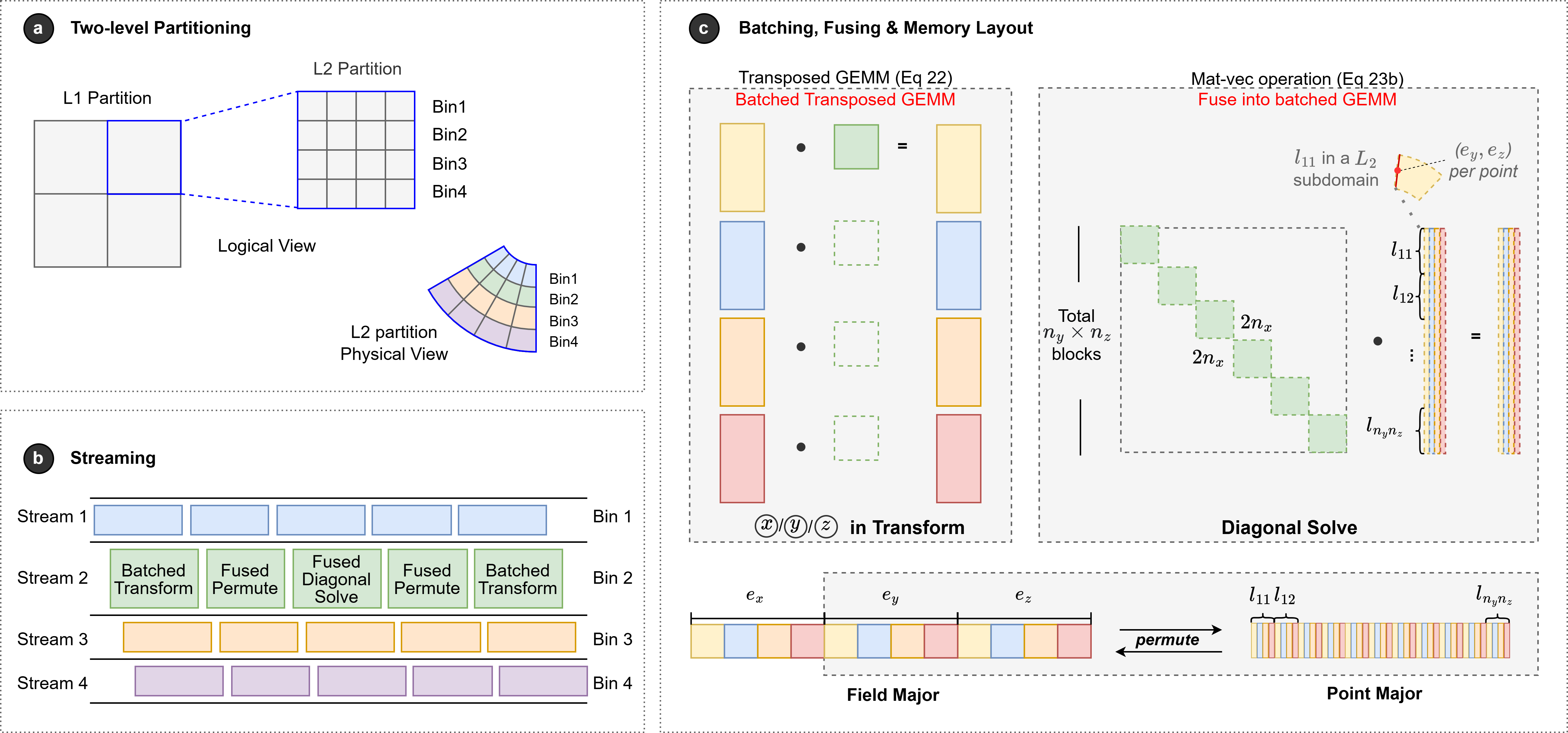}
  \caption{Schematic diagram of HASM for parallel EM simulation. (a) Two-level domain decomposition: \chg{L1 subdomains follow the inter-GPU decomposition, with each assigned to one MPI process and one GPU; each L1 subdomain is further split into L2 subdomains and automatically grouped into bins sharing geometry-dependency of the subdomain solver.
(b) Stream-level execution of an L1 solve: The bins adopt identical color scheme as in (a).
(c) Main operators and data layouts: The green geometry-dependent transformation and block diagonal matrices are shared across all L2 subdomains within each bin. Data exclusive to the L2 subdomains in one bin are color-coded with alternative colors. Left: transforms on field-major variables are batched across subdomains. Right: block-diagonal solves on point-major vectors, fused together and realized by batched GEMM. Note that the fusion operation relies on the custom data layout illustrated at the bottom, where $l_{ij}$ refers to line segments within each L2 subdomain, with $i$ and $j$ corresponding to indices along the $y$ and $z$ direction, respectively.}}
\label{fig:asm}
\vspace{-0.8em}
\end{figure*}

\section{TokaGLINT framework}
\label{secframework}
While the outer Krylov iteration of the TokaGLINT solver is detailed in Section \ref{sec3b}, this section focuses on the HASM preconditioner and the batched multi-subdomain solver implemented on GPU accelerators.

\subsection{Hierarchical ASM}
\label{subsec:hasm}
Figure \ref{fig:asm}(a) illustrates the two-level hierarchical overlapping domain decomposition. While the actual simulation is fully 3D, a 2D schematic is shown for illustrative purposes, \chg{and the detailed overlap regions between neighboring subdomains are omitted for clarity.}
A sector of the global domain is first partitioned into first-level \chg{(L1)} overlapping subdomains, each assigned to a GPU. Each first-level subdomain is then further decomposed into smaller overlapping sub-subdomains \chg{(L2). The L2} subdomains at \chg{same} radial layers in the cylindrical coordinate system \chg{share the same} metric tensors \chg{i.e., the transforms defined in \eqref{eq:newtransform} and \eqref{eq:new inv transform}, and the diagonal solve \eqref{eq:M-in-block-form3-1} and \eqref{eq:M-in-block-form3-2}, and they are grouped into one bin to facilitate operator fusion and batching. The streaming between different bins is illustrated in Figure \ref{fig:asm}(b).}

\chg{As derived in Sec. III.C and illustrated in Figure \ref{fig:asm}(c), two types of expensive operations in the subdomain solve can be identified: the block-diagonal solve \eqref{eq:M-in-block-form3-2} with pre-computed inverse blocks,
which has memory footprint and computational complexity both $O(n^4)$, where n (assuming $n_x=n_y=n_z=n$) denotes the subdomain side length; the forward/backward transforms \eqref{eq:newtransform} and \eqref{eq:new inv transform}, with computational complexity $O(n^4)$. In a single subdomain solve, the transforms are in the form of transposed matrix multiplication, and the block diagonal solve is matrix-vector multiplication. The former is batched within each bin, and the latter is fused into matrix multiplication.}


\chg{The fusion of the block diagonal solve \eqref{eq:M-in-block-form3-2} is the most desirable feature for the L2 domain decomposition and geometry-driven binning. Within a bin, the block solves at the same diagonal-block position can be fused into one GEMM. Consequently, the block-diagonal solves step across L2 subdomains in this bin can be executed as batched GEMM, which is much more efficient than launching multiple separate matrix-vector multiplications. Furthermore, this fusion requires a customized data layout for L2 subdomains within each bin. All of the above designs are illustrated in the right half of Figure \ref{fig:asm}(c).}

\chg{The transforms \eqref{eq:newtransform} and \eqref{eq:new inv transform} natively adopt GEMM form. Nevertheless, small subdomain size $n$ limits single-subdomain performance. By binning subdomains together, we execute them as batched GEMM operations, which coarsens the operation granularity. The reason these operations are not fused together, even though they share identical metric-dependent transform matrices, is that successive transformations along different directions require transposed matrices multiplication. Fusing multiple such operations would demand highly customized low-level implementations, and the performance gains cannot offset the associated development overhead.}

\chg{Binning further enables fusion of additional operations including permutation and L1 communication buffer packing, and so on. Most of them are adopted in our implementation, but since they contribute minor performance gains, we do not elaborate further here.}

To support this grouping strategy efficiently, the solver integrates an automated mechanism for \chg{meta-}data reorganization and mask mapping, handling memory management and boundary condition adaptation.

\subsection{\chg{Discussion}}
\label{subsec:batchedsolver}

\chg{We further discuss alternative strategies corresponding to different features of TokaGLINT.}

\chg{\textbf{Large single-domain solve per GPU.} Different from the large single-domain solve per GPU adopted by many conventional methods, including FlashMP \cite{FlashMP}, our work introduces L2 domain decomposition and geometry-driven binning strategies, effectively reducing the overall computational complexity and fuses, batches, and pipelines matrix transforms, block-diagonal solves, and inter-subdomain data movement across multiple sub-solvers, which large single-domain solves cannot achieve.
}

\begin{chgblock}
We construct a representative example, whose results are presented in Table \ref{tab:l2 ablation}. When solving with a single large subdomain yields poor efficiency (the bottom row), the L2 domain decomposition substantially boosts solving efficiency, with iteration counts staying consistent. The one iteration discrepancy observed in the case for relative tolerance $10^{-10}$ can be attributed to different convergence paths taken to reach the prescribed tolerance.

\begin{table}[htbp]
\centering
\caption{\chg{Representative Benchmarks for L2 Domain Decomposition. Each entry is reported as $T/I$, where $T$ is the total solve time in milliseconds and $I$ is the number of iterations to convergence. Parameters:$\Delta x=\Delta y=\Delta z=1.0$, $\Delta t=10.0$, $r_0=1152$, $l=3$, L1 size $=64\times64\times64$, L1 count $=2\times2\times2$.}}
\label{tab:l2 ablation}
\begin{tabular}{ccccc}
\toprule
  &  & \multicolumn{3}{c}{\chg{Relative tolerance}}\\
  \cmidrule (lr){3-5}
\chg{L2 size} & \chg{L2 count} &\chg{$10^{-8}$} &\chg{$10^{-10}$} &\chg{$10^{-12}$}  \\
 \midrule
\chg{$16\times 16\times16$} & \chg{$4\times 4\times4$} & \chg{49.8/6} & \chg{63.8/8}  & \chg{74.6/9} \\
\chg{$32\times 32\times32$} & \chg{$2\times2 \times2$} & \chg{63.0/6} & \chg{71.7/7}  & \chg{92.1/9} \\
\chg{$64\times 64\times64$} & \chg{$1\times 1\times1$} & \chg{209.0/6} & \chg{238.9/7}   & \chg{302.7/9}\\
\bottomrule
\end{tabular}
\vspace{-0.8em}
\end{table}

\end{chgblock}

 \chg{\textbf{Sparse direct solver.}} Regarding computational cost, the transformation and block-diagonal solve, both exhibit $O(n^4)$ complexity, \chg{which is comparable to the solve-phase cost of sparse direct solvers such as a general purpose sparse LU solve.} \chg{However, sparse triangular solves exhibit inherent data dependencies, and suffer from indirect memory accesses, irregular execution, and unstructured memory access patterns. By contrast, empowered by operator fusion realized through geometry-driven binning of L2 subdomains, the dominant computation within TokaGLINT's solving stage is restructured into batches of concurrent, uniform-sized GEMM operations.}

\chg{\textbf{Why skip the MPS-based approach in \cite{46-Yamazaki2023}?} A key design advantage of TokaGLINT comes from fusing the $O(n^4)$ block-diagonal solves across multiple L2 subdomains, converting many independent matrix-vector multiplications into batched GEMMs. This scheme relies on geometry-driven binning and custom-designed data layouts, which is difficult to achieve without explicit inter-process control of bin partitioning and synchronization. For this reason, the single-process-per-GPU deployment becomes a natural choice.}

\section{Experiments}
All experiments are conducted on China's latest heterogeneous supercomputer supporting both scale-up and scale-out capabilities. Each compute node is equipped with 8 GPUs and two 64-bit CPUs. Each CPU operates at 2.4 GHz with 64 cores, adopts a NUMA-based memory organization, supports eight-channel DDR5-6400 memory, and connects to accelerators via PCIe Gen5. Each GPU integrates 320 SIMD units, with a theoretical peak double-precision (FP64) performance of 32.7 TFLOPS. The GPU is further equipped with 64 GB HBM and supports a theoretical peak memory bandwidth of 1.8 TB/s. Within each node, accelerators are interconnected by a high-speed intra-node accelerator link. Inter-node cluster networking relies on 4$\times$400 Gbps InfiniBand-like, RDMA-capable links. The implementation uses HIP-compatible GPU kernels and GPU-aware MPI for inter-GPU communication. The software stack consists of a GPGPU programming environment compatible with mainstream GPGPU API standards, Clang 17.0.0, and Open MPI 5.0.3.

\chg{In the experiments presented in this section, we use overlap layers $l=3$ for the reason that preliminary tuning showed the best time-to-solution in most cases. The L1 sizes $64^3$--$128^3$ per GPU, are representative of PIC workloads, consistent with VPIC 2.0 using $100^3$ cells per GPU V100~\cite{38-Bird2022} and WarpX reporting tens of millions of cells per GPU in FOM tests~\cite{41-Fedeli2022}. These workloads provide substantial local computation to amortize communication and expose GPU batching efficiency, which should be considered when interpreting the reported scaling efficiency.}

\subsection{Performance}
\label{subsec:speed test}
We compare TokaGLINT against the BiCGStab solver in HIP-enabled HYPRE 2.32. \chg{We use unpreconditioned BiCGStab in HYPRE as the baseline.}
\begin{table}[htbp]
\centering
\caption{Performance comparison with HIP-enabled HYPRE. \chg{Each entry is reported as $T/I$, where $T$ is the total solve time in milliseconds and $I$ is the number of iterations to convergence. Parameters: $\mathrm{tol}_r=10^{-12}$, $\Delta x=\Delta y=\Delta z=1.0$, $\Delta t=8.0$, $r_0=1920$, L1 size $=64\times64\times64$; TokaGLINT: $l=3$, L2 count $=2\times2\times2$.}}
\label{tab:prec_vs_hypre}
\begin{tabular}{lcccc}
\toprule
&  np=8 & np=64 & np=256 &np=512  \\
 \midrule
 HYPRE &   212.38\chg{/125}  &   224.02\chg{/122}   & 235.10\chg{/122} & 257.04\chg{/122}\\
 TokaGLINT &  79.67\chg{/8}  &   81.96\chg{/8}   & 85.09\chg{/8} & 84.79\chg{/8}\\
 \midrule
 Speedup & $2.67\times$&   2.73$\times$  & 2.76$\times$ & 3.03$\times$\\
\bottomrule
\end{tabular}
\end{table}

As shown in Table \ref{tab:prec_vs_hypre}, TokaGLINT achieves a speedup of $2.67\times$ on a single node (8 accelerators), and the speedup ratio further increases as the problem scale expands. This scaling trend is fully consistent with the expected high parallel scalability of the TokaGLINT solver.

\begin{figure}[!h]
  \centering
  \includegraphics[trim=5pt 10pt 5pt 5pt,clip, width=3.2in]{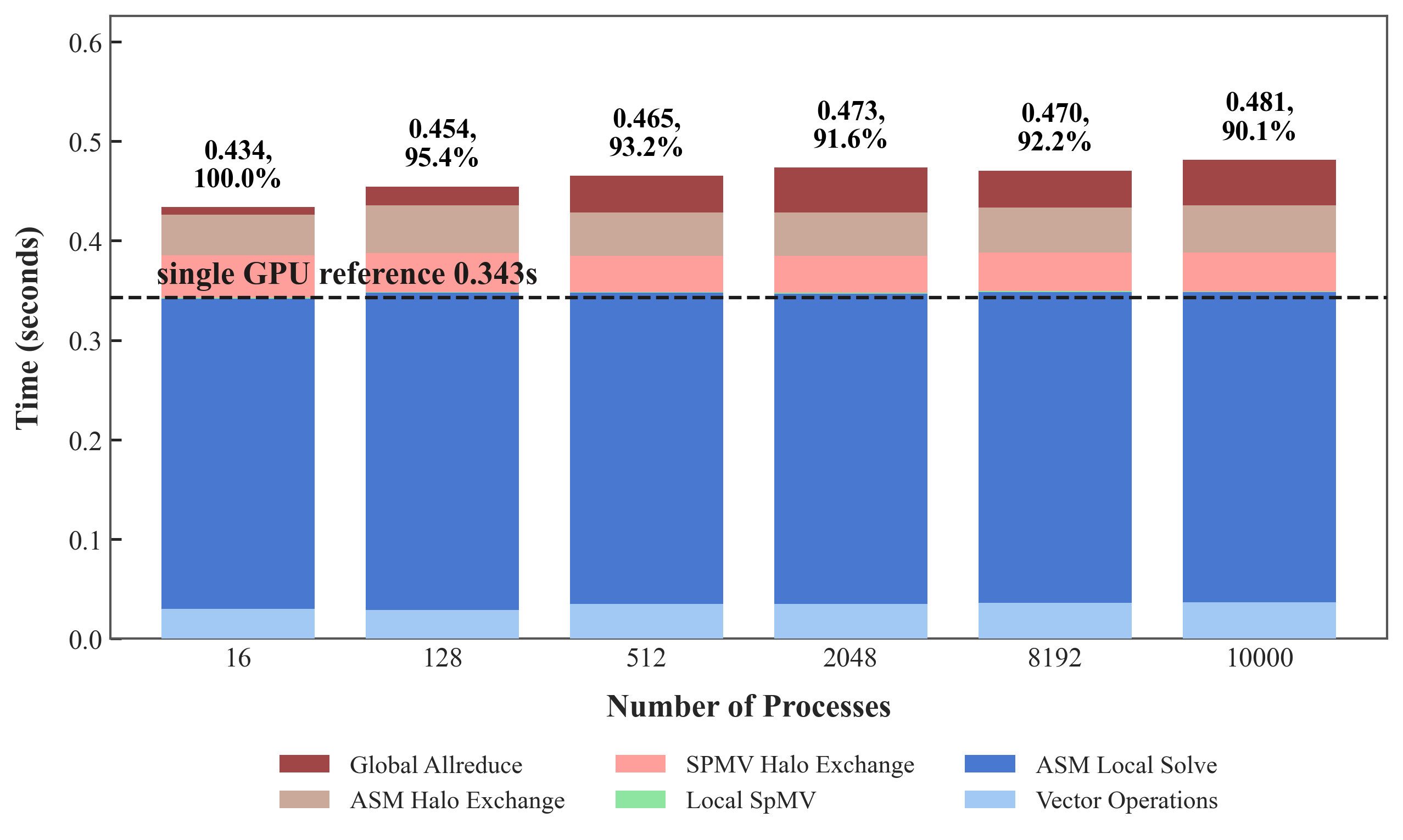}
\caption{\chg{Weak-scaling runtime breakdown of TokaGLINT. Stacked bars show the average time of one linear-system solve to convergence. The two numbers above each bar denote the average solve time in seconds and the parallel efficiency, respectively. The dashed line marks the single-GPU reference case: L1 domain decomposition is absent, so no cross-GPU communication takes place.}}
  \label{fig:weaktest}
\end{figure}

\begin{figure*}[!htbp]
  \centering
  \includegraphics[trim=5pt 10pt 5pt 35pt,clip, width=6.5in]{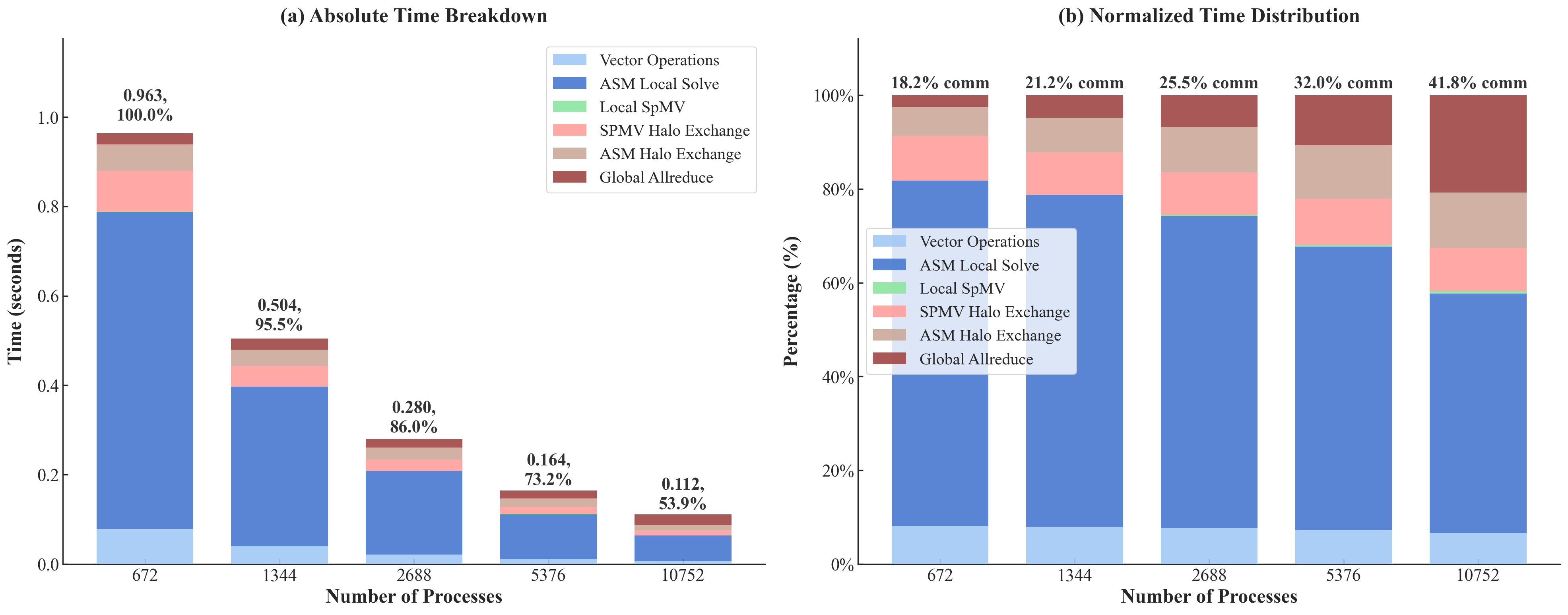}
\caption{\chg{Strong-scaling runtime breakdown of TokaGLINT. The left panel shows the absolute average time of one linear-system solve to convergence, with the two numbers above each bar denoting the average solve time in seconds and the parallel efficiency. The right panel shows the normalized time distribution, with the number above each bar denoting the communication percentage of the total solve time.}}
  \label{fig:strongtest}
\end{figure*}

\subsection{Parallel Scalability}
\label{subsec:weak test}
To assess the scalability of TokaGLINT, we carry out weak- and strong-scaling studies. All reported times in Fig.~\ref{fig:weaktest} and Fig.~\ref{fig:strongtest} are averaged over 100 complete linear-system solves to convergence.

For the weak scaling test (Table~\ref{tab:weakset}), \chgblue{we keep the number of unknowns per accelerator fixed}. \chg{We use the 16-accelerator case (2 nodes) as the scaling baseline because a single-GPU run has no L1 decomposition or inter-GPU communication, and a single-node run mainly exercises direct GPU communication rather than representative multi-node behavior.}

As shown in Fig.~\ref{fig:weaktest}, TokaGLINT achieves an excellent parallel efficiency of \chg{$90.1\%$ when scaling from 16 to 10,000 accelerators ($625\times$)}. These results compellingly confirm TokaGLINT's robust scalability for ultra-large-scale EM simulations.

\begin{table}[ht]
\centering
\caption{Weak-scaling test setup. \chg{Parameters: $\mathrm{tol}_r=10^{-8}$, $\Delta x=\Delta y=\Delta z = 1.0$, $\Delta t=8.0$, $r_0=1920$, $l=3$.}}
\label{tab:weakset}
\begin{tabular}{@{}cccc@{}}
\toprule
Accelerators & Global grid & \chg{L1 count} & \chg{L2 count} \\
\midrule
\chg{16}     & \chg{$512\times256\times256$}   & \chg{$4\times2\times2$} &\chg{$4\times4\times4$} \\
\chg{128}     & \chg{$1024\times512\times512$}   & \chg{$8\times4\times4$} &\chg{$4\times4\times4$} \\
\chg{512}  & \chg{$1024\times1024\times1024$}   & \chg{$8\times8\times8$} &\chg{$4\times4\times4$} \\
2048  & $2048\times2048\times1024$ & $16\times16\times8$  & $4\times4\times4$ \\
8192  & $4096\times2048\times2048$ & $32\times16\times16$ & $4\times4\times4$ \\
10000 & $3200\times2560\times2560$ & $25\times20\times20$ & $4\times4\times4$ \\
\bottomrule
\end{tabular}
\end{table}

For the strong-scaling test, \chgblue{the total problem size is fixed, as listed in Table~\ref{tab:strongset}}. TokaGLINT reaches the relative tolerance in 6 iterations for all tested parallel configurations. As shown in Fig.~\ref{fig:strongtest}, when scaling from 672 to 10,752 accelerators, TokaGLINT maintains a parallel efficiency of $53.9\%$.

\begin{table}[ht]
\centering
\caption{Strong-scaling test setup. \chg{Parameters: $\mathrm{tol}_r=10^{-7}$, $\Delta x=\Delta y=\Delta z = 1.0$, $\Delta t=8.0$, $r_0=1920$, $l=3$.}}
\label{tab:strongset}
\begin{tabular}{@{}cccc@{}}
\toprule
Accelerators & Global grid & \chg{L1 count} & \chg{L2 count} \\
\midrule
672   & $2048\times1536\times1792$ & $8\times6\times14$   & $8\times8\times4$ \\
1344  & $2048\times1536\times1792$ & $8\times12\times14$  & $8\times4\times4$ \\
2688  & $2048\times1536\times1792$ & $16\times12\times14$ & $4\times4\times4$ \\
5376  & $2048\times1536\times1792$ & $16\times12\times28$ & $8\times8\times4$ \\
10752 & $2048\times1536\times1792$ & $16\times24\times28$ & $8\times4\times4$ \\
\bottomrule
\end{tabular}
\end{table}

Thanks to the communication-hiding BiCGStab solver, global Allreduce operations are overlapped with preconditioning and SpMV, effectively reducing communication overhead. As shown in Fig.~\ref{fig:strongtest} (right), the fractions of SpMV and ASM halo exchange rise with parallel scaling, while Allreduce increases more significantly, though the overall communication ratio remains below 50\%. Owing to the high computational cost of subdomain solvers, the algorithm maintains good strong-scaling efficiency. The observed efficiency loss mainly comes from the growing Allreduce proportion and reduced local computation time. At large core counts, lower per-accelerator computational intensity limits the batched local solver from reaching peak performance. Allreduce suffers the largest performance degradation, followed by ASM halo communication, whose data volume is three times that of SpMV halo exchange.

\subsection{Simulation}
\begin{figure*}[!htbp]
\centering
\includegraphics[trim=5pt 10pt 5pt 5pt,clip, width=6in]{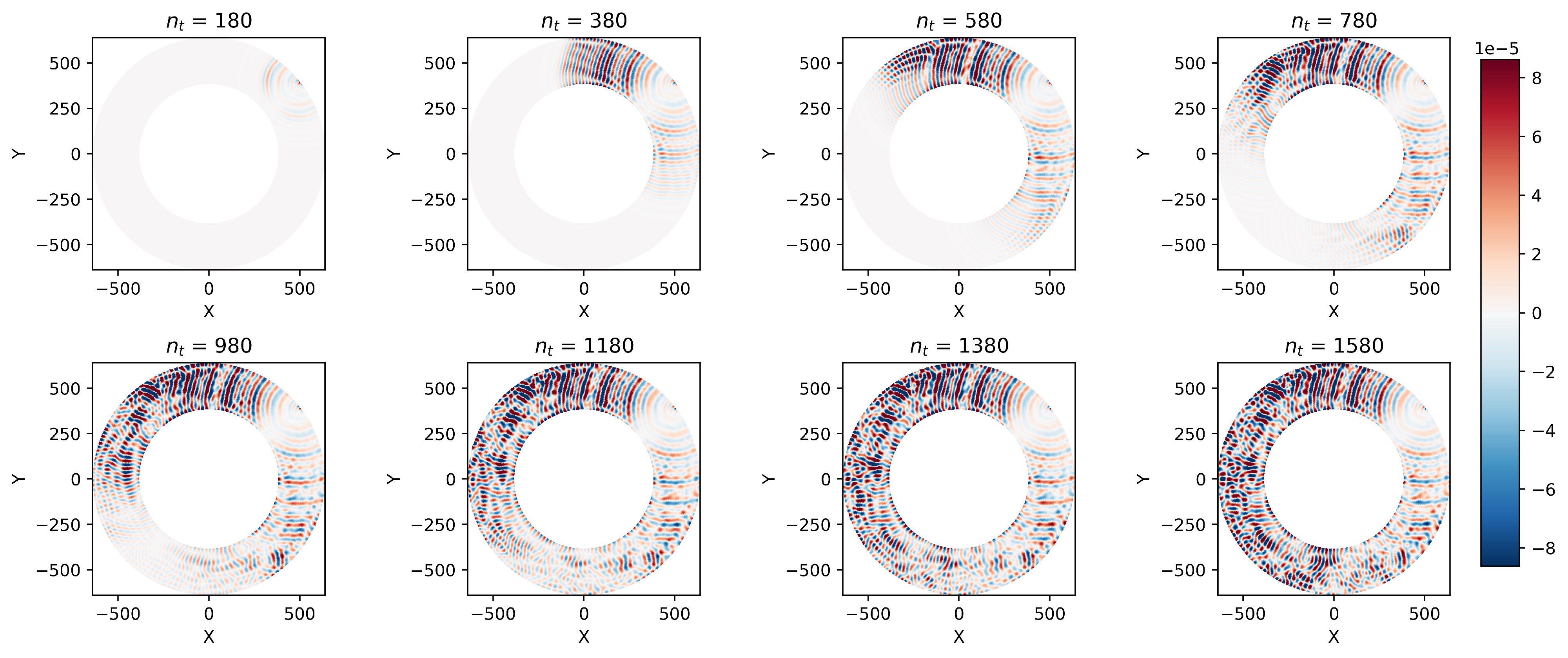}
\caption{\chg{Wave-propagation validation of the CN-FDTD field solver in SymPIC using TokaGLINT. Wave source is located at $(X\approx 519\Delta l, Y\approx 375\Delta l)$. The panels show time snapshots of the Ey value on the z-midplane, visualized in Cartesian coordinates after mapping the logical cylindrical grid $(x,y,z)$ to $(X,Y,Z)$.}}
\label{allB2}
\end{figure*}
\label{subsec:real case test}
\chgblue{We next validate the integration of TokaGLINT into the SymPIC EM field solve through a wave-propagation test and a long-time energy-conservation test. We first test the correctness of the field solver using a wave-excitation case. The simulation} domain is
\begin{gather*}
n_x=n_y/4=n_z=256,  \\[2pt]
r_0=384\Delta l, \Delta x=\Delta z=\Delta l, \Delta y=2\pi r_0\Delta l/n_y~.
\end{gather*}
PEC boundaries are adopted for $x$ and $z$ directions, while periodic boundary condition
is chosen for $y$ direction. Wave source is located at $x=n_x\Delta x, y=102\Delta y, z=n_z\Delta z/2$ and the frequency is $\omega=0.2\mathrm{c}/\Delta l$. Time step is set to $\Delta t=2\Delta l/c$ and total number of time steps is $n_t=1600$.
Evolution of electric field is shown in Fig.~\ref{allB2}, which clearly \chg{shows} the propagation process of the EM wave inside the toroidal domain.
For visualization, results computed in the logical cylindrical coordinate system $(x,y,z)$ are mapped to Cartesian coordinates $(X,Y,Z)$ and plotted on the slice $Z=128$. The coordinate transformation is defined as follows:
\begin{gather*}
X=\left(i\Delta x+r_0 \right) \cos(j \Delta y/r_0)~,  \\[2pt]
Y=\left(i\Delta x+r_0 \right) \sin(j \Delta y/r_0)~, \quad Z=z~.
\end{gather*}

A key merit of adopting symplectic structure-preserving PIC scheme is that the truncation errors of fundamental system invariants such as the total energy remain uniformly bounded over long simulation time intervals. To verify this property, we performed a three-dimensional magnetized electron toroidal plasma simulation with the parameters specified below.
\[
\begin{gathered}
\lambda_d = 1.85 \times 10^{-3} \Delta l, \\
\omega_{pe,0} \Delta l / \mathrm{c} = 2.82 \times 10^{-1},\quad \omega_{ce,0} \Delta l / \mathrm{c} = 1.69 \times 10^{-1}, \\
\boldsymbol{B} = B_0 \vec{\boldsymbol{j}}, \\
x_0=1920\Delta l, \quad n_x = 2n_y = n_z = 128,\\
\Delta x = 1.0 \Delta l, \quad \Delta y = 1.2 \Delta l, \quad \Delta z = 1.3 \Delta l, \\
\Delta t c / \Delta l = 2.5,
\end{gathered}
\]
\noindent where $\omega_{pe,0} = \sqrt{\frac{n_0 q_e^2}{\varepsilon_0 m_e}}$, $v_{te}$, $\lambda_d = v_{te} / \omega_{pe}$ and $\omega_{ce} = \frac{q_e B_0}{m_e}$ are the plasma frequency, thermal speed, Debye length and cyclone frequency of electrons, $n_0$ is the reference density,
density distribution of electrons is
\[ n_{e,i}(x,y,z) = n_0 e^{-R^2 / R_0^2}\]
where $R=\sqrt{(x-64)^2+(z-64)^2},R_0=7.1$,
$\mathrm{c}$ is the speed of light in the vacuum. The total number of simulation \chg{time steps} is $n_t =1.0 \times 10^6$, which means $\Delta t n_t \omega_{pe} = 7.05 \times 10^5$.
where $m_e$,  $q_e$,  $v_{te}$ are mass, charge, thermal speed of electrons, respectively. The external magnetic field is $ \boldsymbol{B} = B_0 \vec{\boldsymbol{j}}$, and the number of sampling particles per grid at $R=0$ is set to 56.

\begin{figure}[!htbp]
    \centering
    \includegraphics[ trim=5pt 10pt 5pt 5pt,clip, width=3.1in]{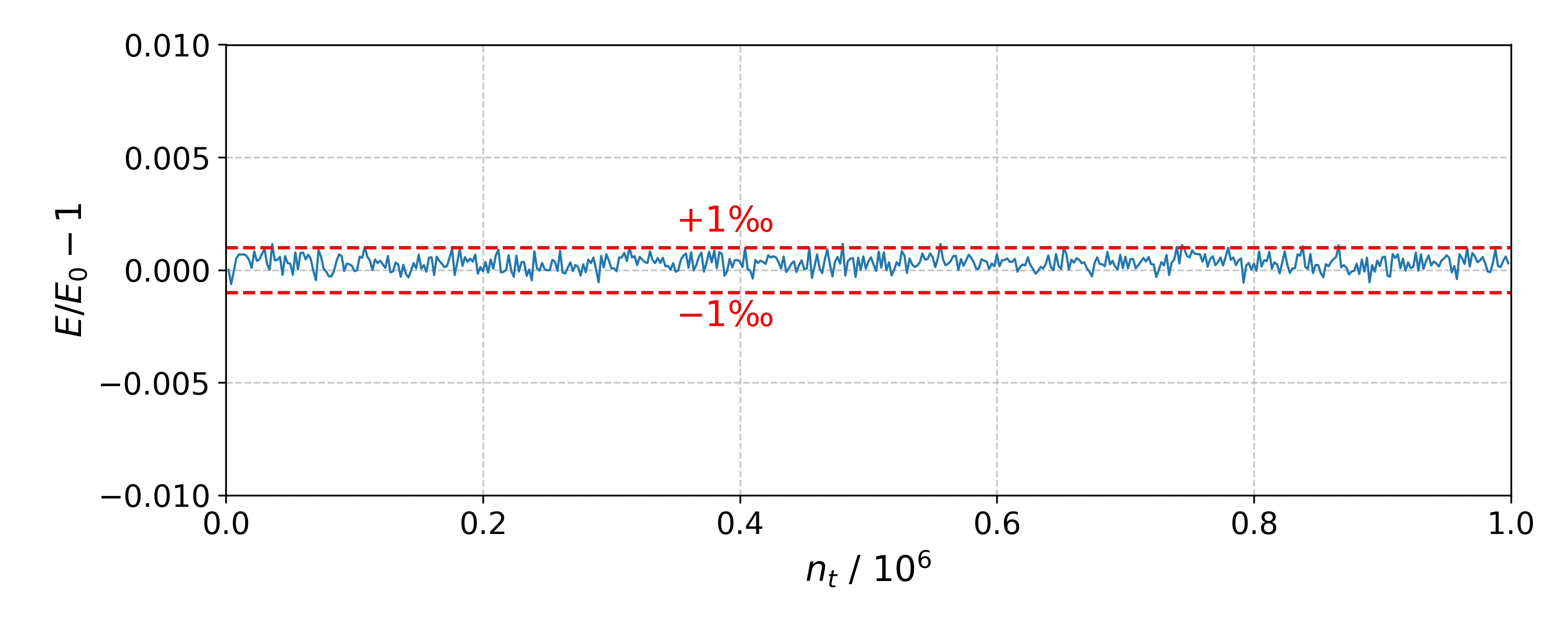}
\caption{\chg{Long-time total-energy variation of a magnetized toroidal electron plasma simulated by the field-implicit symplectic PIC scheme in SymPIC using TokaGLINT.}}
    \label{fig:energy_evolution}
    \vspace{-0.8em}
\end{figure}

The total energy is recorded every 2000 \chg{time steps}, which is shown in Fig.~\ref{fig:energy_evolution}. As the initial field is far from equilibrium, the energy at the second output step ($n_t=2000$) is taken as the baseline $E_0$. The internal energy $E$ is found to vary within $\pm 1\text{\textperthousand}$ of $E_0$ for the rest of the simulation. It is clear that the total energy is well conserved, which verifies the advantage of the present field-implicit symplectic PIC scheme for cylindrical coordinates.

\section{Conclusion}
In this work, we propose TokaGLINT, a highly scalable, hardware-aware linear solver designed to address the computational bottlenecks of the CN-FDTD discretization of Maxwell's equations in curvilinear coordinates on modern heterogeneous GPU platforms, as part of the upgrade to the EM solver of the large-scale parallel PIC simulation software SymPIC. Built upon a hierarchical domain decomposition framework coupled with tensor-structured fast subdomain solvers tailored for curvilinear coordinate systems, this composite preconditioner effectively overcomes the scalability challenges of linear system solution on GPU clusters, delivering substantial speedup and excellent parallel scalability while preserving rigorous numerical fidelity for large-scale tokamak simulations.

Extensive numerical experiments thoroughly verify the effectiveness and competitiveness of the proposed method. Compared with the widely adopted general-purpose solver library HYPRE, TokaGLINT achieves a performance improvement of $2.67\times$ on a single node. At massively parallel scales, the solver maintains a weak scaling efficiency of \chg{90.1\%} and a strong scaling efficiency of 53.9\% across more than 10,000 GPUs. Moreover, its practical deployment within the SymPIC code demonstrates its capability to enable highly efficient implicit EM field solutions under realistic physical scenarios.

Future work will aim to broaden the applicability to more diverse physical scenarios and multiphysics simulation workflows.  In particular, we will investigate fully implicit formulations that treat both electromagnetic field evolution and particle kinetics within a consistent implicit framework.

\section*{Acknowledgments}
We sincerely thank our anonymous SC reviewers for their valuable comments. This work was supported by the National Key Research and Development Program of China (Grant No. 2025YFB3003403) and the Strategic Priority Research Program of Chinese Academy of Sciences (Grant No. XDB0500101).


\balance

\vspace{12pt}

\end{document}